\documentstyle[a4,12pt]{article}
\begin{document}
\begin{titlepage} \vspace{0.2in} \begin{flushright}
MITH-97/13 \\ \end{flushright} \vspace*{1.5cm}
\begin{center} {\LARGE \bf  Coherent Electrodynamic Processes in Solid 
Molecular Hydrogen\\} \vspace*{0.8cm}
{\bf M.~Buzzacchi, E.~Del Giudice, G.~Preparata, S.S.~Xue$^{(a)}$}\\ \vspace*{1cm}
INFN - Section of Milan, Via Celoria 16, Milan, Italy\\
Dipartimento di Fisica dell'Universit\`a di Milano\\ \vspace*{1.8cm}

{\bf   Abstract  \\ } \end{center} \indent
\baselineskip=18pt
The many experimental puzzles of solid hydrogen, that seem to defy the 
generally accepted view of condensed matter, are shown to be a rather 
straightforward consequence of a new approach to condensed matter 
physics, that takes into account the coherent electrodynamic 
processes, predicted by Quantum ElectroDynamics ($QED$), that have 
been so far overlooked. As a consequence, a revision of the present 
view, based on electrostatic forces only, is called for.
\baselineskip=12pt 
\vfill \begin{flushleft}  November 1997   \vspace*{3cm} \\
\noindent{\rule[-.3cm]{5cm}{.02cm}} \\
\vspace*{0.2cm} \hspace*{0.5cm} ${}^{a)}$ 
E-mail address: xue@mi.infn.it\end{flushleft} \end{titlepage}
\baselineskip=12pt
 
Much as the hydrogen atom has been the paradigm-problem of Atomic 
Physics, it would seem that solid hydrogen should have the same role 
in solid state physics, a simple and essential testing ground of the 
prevailing theoretical ideas on condensed matter. And yet, solid 
hydrogen has become over the years an embarassing puzzle, a source of 
remarkable discoveries that have found the generally accepted 
paradigm, that somewhat impressionistically we may describe as the 
Electrostatic Meccano \footnote{According to the generally accepted 
view, the interactions among the atomic/molecular systems of condensed 
matter have an electrostatic nature, hence a short-range character, 
just like the mechanical forces that hold together the pieces of a 
Meccano.}, quite off-balance.
Indeed, there are two major aspects of solid hydrogen phenomenology 
that, in our opinion, expose the inadequacy of the present view: at low 
pressure the observed equation of state defies all efforts based on 
"reasonable" two-body potentials, and appears to require the action of 
"many-body forces" \cite{mh94}, while at high pressure metallic 
hydrogen, predicted by the classic work of Wigner and Huntington 
\cite{wh} to occur at pressures of 25 GPa, at pressures ten times 
larger still eludes detection. On the other hand, for $P\simeq$ 150 
GPa a new remarkable phase of molecular hydrogen, namely solid 
hydrogen III, has been discovered about ten years ago, characterized 
by intense infrared absorption and a sharp discontinuity in the vibron 
frequencies \cite{mh88}, \cite{ls89}.

In this Letter we wish to report a few results on the physics of solid 
molecular hydrogen of a new approach to condensed matter which, based 
on $QED$, takes finally into account the non-negligible collective 
electrodynamic actions that the elementary systems (atoms and 
molecules) exert upon each other at high enough densities and low 
enough temperatures. As the basic theory, as well as a few relevant 
applications to the physics of condensed matter have been published in 
a recent, readily accessible book \cite{gp95}, here we shall refrain 
to review the theory \footnote{We believe it should also be of some 
interest to the reader to consult a recent article \cite{enz97} by 
C.P. Enz, where the basic theoretical ideas of $QED$ coherence are 
critically analysed.}, and only mention the two theoretical 
developments which, as we will show, have an important bearing upon 
the physics of solid molecular hydrogen, namely the application of the 
$QED$ coherent interaction to the solidification of $^{4}$He 
\cite{gp95}, \cite{demp} and to the dynamics and thermodynamics of 
liquid water \cite{gp95}, \cite{abdp}.

Let us then address first the problem of solidification of molecular 
hydrogen at low pressures. We shall follow London \cite{london} in 
modeling the solid as an ensemble of cages where each molecule 
"rattles", confined by its nearest neighbours. As we are not 
interested in a detailed calculation of the spatial structure of the 
solid, but only on the collective electrodynamic interactions, that 
have been so far neglected, we shall model the cage as a spherical 
cavity \cite{demp} of radius $R=r-r^{*}$, where $r$ is the average 
intermolecular distance and $r^{*}$ is the radius of the co-volume, 
that we evaluate by imposing that at $r^{*}$ the total energy, kinetic 
plus Lennard-Jones, be zero. For $H_2$ and $D_2$ we obtain 
$r^{*}=2.79$ $\AA$  and $2.80$ $\AA$  respectively. As explained in 
Refs.\cite{gp95}, \cite{demp}, the main idea is that due to its repulsive 
interaction with the nearest neighbours, the walls of its cage, each 
molecule, slightly violating the Born-Oppenheimer approximation, 
develops a weak electric dipole, through which it gets coupled to the 
quantized modes of the electromagnetic field.
By means of these simple considerations we may look at our molecules as 
two-level systems, "oscillating" between the $s$-wave ground state 
and the $p$-wave first-excited state, whose energies are
\begin{equation}
E_0=\frac{\pi^2}{2mR^2}~,~~~~~E_1=\frac{(4.49)^2}{2mR^2}
\end{equation}
respectively \footnote{We are using here the "natural units", where 
$\hbar=c=k_{B}=1$}, resonantly coupled to the electromagnetic modes of 
frequency
\begin{equation}
\omega=E_1-E_0=\frac{10.29}{2m(r-r^{*})^2}.
\end{equation}
In order to complete the quantitative characterization of the 
two-level system we must estimate the size of its electric dipole, 
which determines the strength of its coupling to the radiation field.
As argued in Refs. \cite{gp95}, \cite{demp}, a reliable estimate of 
the dipole $D(r)=2ed(r)$ can be obtained by equating the repulsive 
potential of two dipoles at the distance $r$ with the repulsive part 
of the Lennard-Jones potential, i.e. by setting
\begin{equation}
\Delta\left(\frac{R_m}{r}\right)^{12}=4\alpha\frac{d(r)^{2}}{r^3},
\end{equation}
where $\Delta$=36.7 K and $R_m$=3.32 $\AA$  for $H_2$ (35.2 K and 3.31 
$\AA$  in $D_2$) as determined by the 
second virial coefficient of the gas phase \cite{michels}, and 
$\alpha=\frac{1}{137}$ is the fine structure constant.

Once we know $\omega(r)$, and the effective coupling constant given by
\begin{equation}
g(r)=\left(\frac{4\pi}{3}\right)^{1/2}\frac{2ed(r)}{\omega(r)^{1/2}}\left(
\frac{N}{V}\right)^{1/2},
\end{equation}
the coherent many-body electrodynamic interaction is fully determined 
\cite{gp95}, \cite{demp}, and we find that for $P$~=~0 and $T$~=~0, at 
the density of the solid  
\begin{equation}
g(r=3.79 A)=1.72 > g_{C}=\left(\frac{16}{27}\right)^{1/2}\simeq 0.77,
\end{equation}
Similarly for deuterium we find $g(r=3.71 \AA)$=2.32, that is we are 
in the regime of strong coherence for both isotopes. 
In Figure 1 we report the energetics of both 
solid hydrogen and deuterium as a function of the intermolecular 
distance $r$. Please note that $\omega(r)\rightarrow\infty$ as 
$r\rightarrow r^{*}$ and Eq.(4) 
predicts that $g(r)<g_{C}$, thus for such values 
of $r$ the energy gain from the long-range coherent interaction 
disappears. For the zero-pressure binding energies of a molecule in the 
lattice, we obtain the values $E$=~-~83~K for H$_2$ (exp.~ -~85.5~K 
\cite{silvera}) 
and $E$=~-~127~K for D$_2$ (exp.~-132~K \cite{silvera}). 
It should also be perfectly clear that for this mechanism 
to be operative one needs a well defined $\omega(r)$, a requirement 
that only a crystalline structure can meet.

Without much ado, and in an obviously perfectible way, we may extend 
our calculation for $T\neq$0. In Ref. \cite{gp95} it is shown that the 
disordering effect of thermal fluctuations produces a two-fluid 
system, comprising an ordered solid in patches of the size of the 
Coherence Domain (CD), of radius $R_{CD}\simeq \frac{\pi}{\omega(r)}$, 
and a disordered fluid that resides in the interstices. The 
$T$-dependence of the coherent, ordered fraction is given by (Figure 
2):
\begin{equation}
\rho_{C}(T)=1-\frac{4\pi}{V_{CD}}\int_{0}^{R_{CD}}r^{2} 
exp\left(-\frac{\delta(r)}{T}\right)dr,
\end{equation}
where $V_{CD}$ is the volume of the CD, and $\delta(r)$ is a well 
defined $r$-dependent gap \cite{gp95}, that arises from the coherent 
electrodynamic process we have described above. Thus, for the free 
energy per molecule in the solid we may write:
\begin{equation}
F_{solid}(T)=\rho_{C}(T)\delta_{0}+\left(1-\rho_{C}(T)\right)\left[
-T\ln\left(e\left(\frac{V}{N}\right)_
{solid}Z_{rot}\right)\right]+E_{e.s.,liquid}+\delta_{e.s.},
\end{equation}
where $Z_{rot}$ is the rotational partition function of the molecule, and 
$\delta_{e.s.}$ is the "electrostatic" gap between the solid and the 
liquid phase, which in the mean field approximation, by use of the 
pair-potential and the values of the molar volumes in the liquid 
$V_{L}=26.1$ cm$^{3}$ and in the solid $V_{S}=23.15$ cm$^{3}$, turns out to 
equal $\delta_{e.s.}=E_{e.s.,solid}-E_{e.s.,liquid}\simeq -4$ K.
In a similar fashion we have for the liquid free energy per molecule:
\begin{equation}
F_{liquid}(T)=-T\ln\left[e\left(\frac{V}{N}\right)_{liquid}
Z_{trans}Z_{rot}\right]+E_{e.s.,liquid}
\end{equation}
where $Z_{trans}$ is the translational partition function. In Figure 3 we 
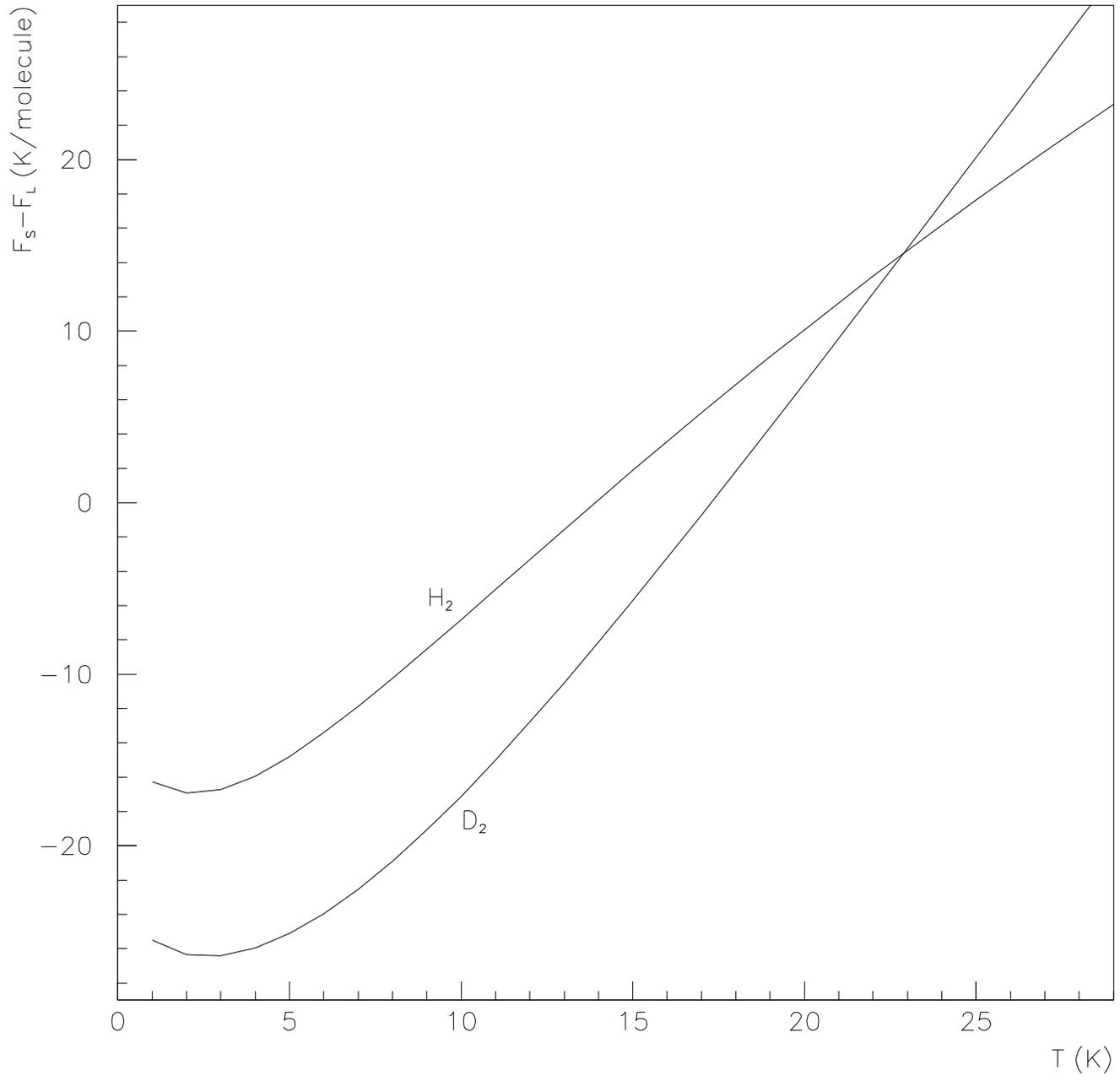
have plotted the free energy difference per molecule between the solid 
and the liquid phase, as a function of temperature for $P$=0. We 
see that the two curves cross zero at $T$=13.5 K for $H_2$ and 
$T$=17 K for $D_2$, in adequate agreement with the observed 
solidification temperatures $T_{SL}$=13.96 K of $H_2$ and $T_{SL}$=18 
K of $D_2$ respectively.

For finite T, as the pressure increases, London's spherical cage 
ceases to be a good approximation, as we can easily understand from 
the vanishing of the gap when $r \rightarrow r^{*}$ (see Fig.~1).  Thus 
at high pressure we expect  the precise spatial structure and the 
rotational state of the molecules to realize a new "cage 
configuration", whose energetic advantage will possibly lead, as 
observed, to a transition to the phase II.

Leaving the complex and fascinating subject of phase II at this 
qualitative level of understanding, we wish now to address the 
extremely intriguing problem of phase III, reached at pressures of  
about 150 GPa. As well known, the signal of this phase transition is 
a gigantic increase in the absorbance of the infrared (IR) vibron mode 
and a sharp discontinuity $\Delta\nu$ in the frequencies of both IR 
and Raman vibrons. Both facts have been, in our opinion, correctly 
interpreted in a recent paper \cite{ae97}, as arising from a sudden 
change in the electronic configuration of the $H_2$ ($D_2$) molecule, 
which thus acquires a permanent electric dipole.  However, we think 
the mechanism of dielectric instability, proposed in the above work to 
account for this most interesting phase transition, hardly believable.
We shall now show that QED coherence has something very relevant to 
say also for phase III. Indeed, even though the electromagnetic 
couplings of the molecular ground state to the excited states ($\Delta 
E\geq 10$ eV) is known to be rather small, like it happens in the 
$H_{2}O$ molecule \cite{abdp}, the access to a coherent ground state 
is made possible by a photon "mass term" inside the solid given by: 
\footnote{See for instance Ref. \cite{gp95}, pages 50-53 and 195-205}
\begin{equation}
\mu_{\gamma,\vec{k}}^{2}=-e^{2}\left(\frac{N}{V}\right)\sum_{n}
\frac{\omega_{\vec{k}}^{2}\left(E_{n}-E_{\alpha}\right)}{\left(E_{n}-E_{\alpha}
\right)^{2}-\omega_{\vec{k}}^{2}}\sum_{i,j=1,2,3}\sum_{r=1,2}\langle\alpha|
\vec{\epsilon}_{\vec{k},r}^{*}\cdot\vec{x}_{j}|n\rangle\langle 
n|\vec{\epsilon}_{\vec{k},r}\cdot\vec{x}_{i}|\alpha\rangle,
\end{equation} 
where $\vec{k}$ refers to the electromagnetic field mode with momentum 
$\vec{k}$, $\omega_{\vec{k}}$ is the energy of the mode 
($\omega_{\vec{k}}=|\vec{k}|$ in free space) and $|n\rangle$ is a 
complete set of molecular states of energy $E_{n}$. By performing an 
analysis totally similar (but for the different experimental inputs: 
oscillator strengths and energy spectrum) to the one carried out for 
water \cite{gp95}, \cite{abdp}, we obtain the following results:\\
\begin{enumerate} 
\begin{itemize}

\item the threshold for the coherent electrodynamic process that 
rearranges the electronic configuration of the $H_2$ ($D_2$) molecule 
occurs at
\begin{equation}
\frac{\rho}{\rho_0}\simeq8
\end{equation}
to be compared  with $\left(\frac{\rho}{\rho_0}\right)_{exp}\simeq 9$

\item in the new configuration the molecules are in a state
\begin{equation}
|C\rangle=\cos\theta |0\rangle+ \sin\theta 
e^{-i\omega_{r}t}|\Pi_{u}\rangle,
\end{equation}
with 
\begin{equation}
\sin \theta = 0.28,
\end{equation}
a superposition of the ground state $|0\rangle$ and the molecular 
configuration $|\Pi_{u}\rangle$, whose energy in the gaseous state is 
$E(\Pi_{u})$=12.4 eV;

\item the energy gain per molecule, the gap $\delta_{e}$, turns out to 
be rather 
small:
\begin{equation}
\delta_{e}\simeq~-~0.10~eV.
\end{equation}
\end{itemize}
\end{enumerate}

Estimating the electric dipole of the $|\Pi_{u}\rangle$ state by 
analogy with the HCl molecule \cite{herzberg}
\begin{equation}
\frac{D}{e}\simeq2\cdot10^{-9} cm
\end{equation}
yields an oscillator strength for the IR vibron
\begin{equation}
f_{th}\simeq\frac{2}{3}m_{e}\omega_{vibron}\sin^{4}\theta\left(\frac{D}{e}\right
)^{2}=
1.13\cdot10^{-5},
\end{equation}
to be compared with $f_{exp}\simeq 1.2\cdot10^{-5}$ \cite{mh94}.\\

With the above informations we may also evaluate the frequency shifts 
$\Delta\nu$ of the vibrons in Phase III. Indeed, our model predicts 
that, as the threshold for electronic 
coherence is reached, the vibrational dynamics of the protons will 
undergo a discontinuous change. In fact, when the electronic 
wave function acquires a small but finite admixture of 
the dipolar configuration $|\Pi_{u}\rangle$, the protons no longer 
oscillate in the previous potential well. Quantitatively, we took the 
experimentally known frequencies of the intramolecular stretching 
modes in the gaseous phase for the ground and 
first-excited electronic levels , in which the vibron 
has the values $\nu$=4395 (3118) cm$^{-1}$ and $\nu^{*}$=2442 
(1736) cm$^{-1}$ in H$_2$ (D$_2$) respectively 
\cite{herzberg}, and the value of the "mixing angle" $\theta\simeq$0.29.\\

In the coherent electronic ground state (Eq. 11) which marks phase III, 
the old vibron is replaced by a new vibron, whose frequency is:
\begin{equation}
\nu_{III}=\nu\cos^{2}\theta+\nu^{*}\sin^{2}\theta
\end{equation}
i.e. 4235 cm$^{-1}$ in H$_2$ and 3005 cm$^{-1}$, yielding downward 
shifts $\Delta\nu=160$ cm$^{-1}$ in H$_2$ and 113 cm$^{-1}$ in 
D$_2$, only somewhat larger than those observed by Mao and Hemley 
\cite{mh88} at finite (T=77 K) temperature, i.e. about 103 cm$^{-1}$ 
for deuterium.  
In view of our neglect of the effect of pressure and temperature 
on the vibron frequencies, we regard this result as quite satisfactory.

In Figures 4a,~4b  we present our results for the equation of state 
of solid hydrogen compared with recent experimental data 
\cite{loubeyre}. It provides better agreement than other calculations 
based on a purely electrostatic approach \cite{duffy}.

In conclusion, we believe we have provided sufficient evidence to 
recommend a serious and rational reconsideration of the prevailing 
views of condensed matter, especially in the light that the approach, 
which we have shown to account effortlessly for many puzzles of 
solid hydrogen, encompasses {\it all} the $QED$ interaction 
mechanisms, not merely the electrostatic ones. 

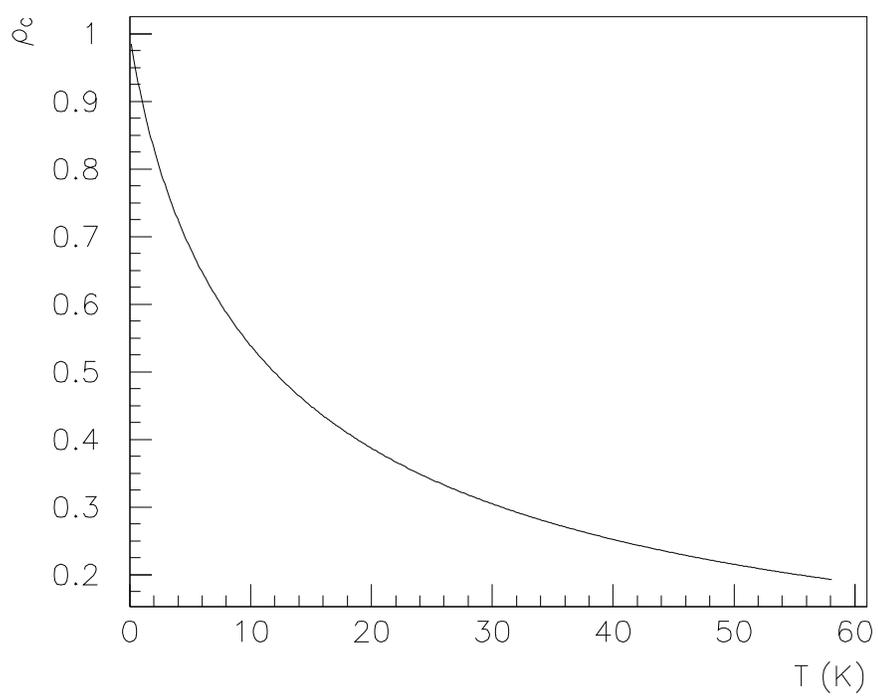
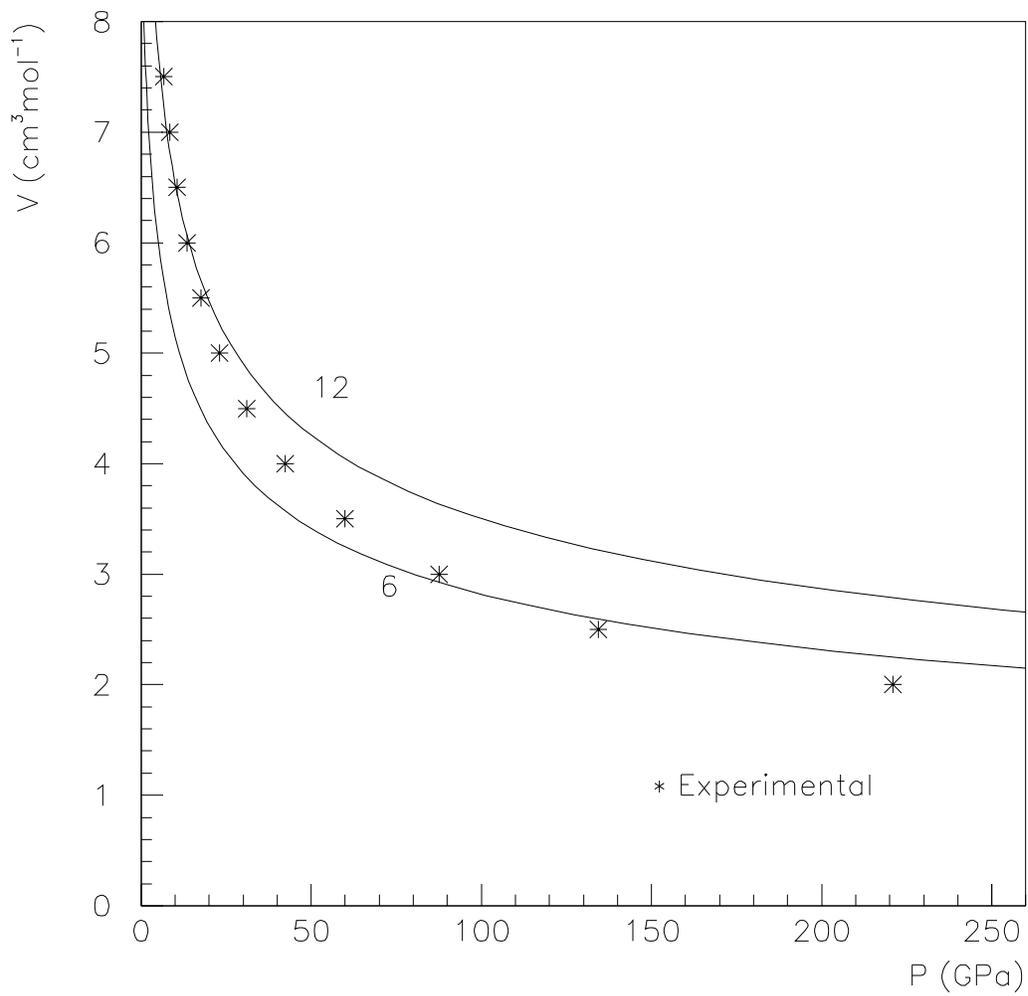
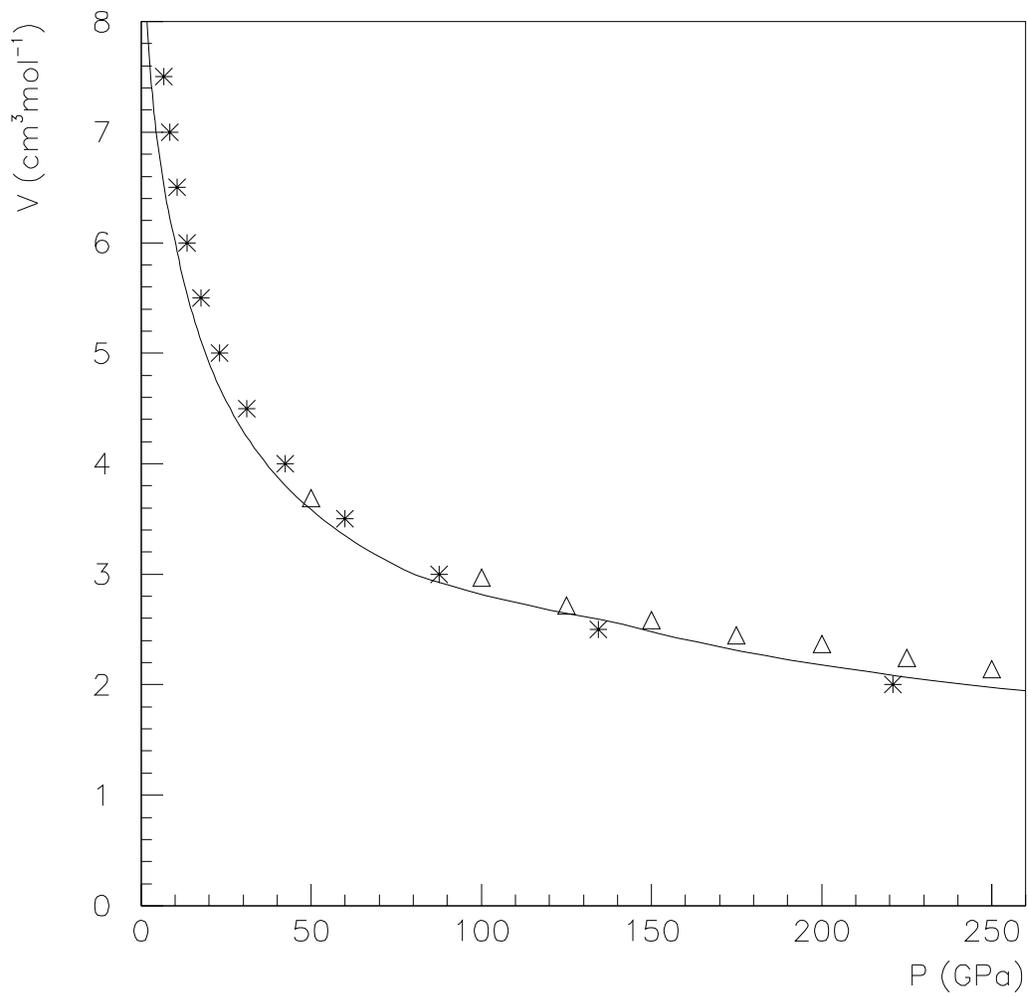
\end{document}